\documentclass[twocolumn,showpacs,aps,prb]{revtex4}%
\usepackage{amsmath}
\usepackage{amsfonts}
\usepackage{amssymb}
\usepackage{graphicx}
\usepackage{float}
\usepackage{subfigure}
\usepackage[colorlinks,  linkcolor=blue, anchorcolor=blue,citecolor=blue,
urlcolor=blue ]{hyperref}
\usepackage{txfonts}%
\usepackage{dcolumn}
\usepackage{bm}

\makeatletter

\newcommand{\Rmnum}[1]{\expandafter\@slowromancap\romannumeral #1@}
\makeatother
\begin{document}
\title{Electronic and magneto-optical properties of monolayer phosphorene quantum dots}
\author{Rui Zhang$^{1}$, X.Y. Zhou$^1$, D. Zhang$^1$, W. K. Lou$^1$, F. Zhai$^2$, Kai Chang$^1$\footnote{kchang@semi.ac.cn}}
\affiliation{$^{1}$SKLSM, Institute of Semiconductors, Chinese Academy of Sciences, P.O.
Box 912, 100083 Beijing, China }
\affiliation{$^{2}$Department of Physics, Zhejiang Normal University, Jinhua 321004, China. }

\begin{abstract}
We theoretically investigate the electronic and magneto-optical properties of rectangular, hexangular, and triangular monolayer phosphorene quantum dots (MPQDs) utilizing the tight-binding method. The electronic states, density of states, electronic density distribution, and Laudau levels as well as the optical absorption spectrum are calculated numerically. Our calculations show that: (1) edge states appear in the band gap in all kinds of MPQDs regardless of their shapes and edge configurations due to the anisotropic electron hopping in monolayer phosphorene (MLP). Electrons in any edge state appear only in the armchair direction of the dot boundary, which is distinct from that in graphene quantum dots; (2) the magnetic levels of MPQDs exhibit a Hofstadter-butterfly spectrum and approach the Landau levels of MLP as the magnetic field increases . A "flat band" appears in the magneto-energy spectrum which is totally different from that of MLP;
(3) the electronic and optical properties can be tuned by the dot size, the types of boundary edges and the external magnetic field.
\end{abstract}
\maketitle
\section{INTRODUCTION}
Since its discovery in 2004\cite{K.S.Novoselov}, graphene has become a major issue of theoretical and experimental study in condensed-matter physics\cite{K.S.Novoselov,A.H.CastroNeto,A.K.Geim1}. However, graphene suffers from limitations for practical applications due to the absence of a band gap. The discovery of graphene evokes great interest and triggers intensive investigation on two-dimensional (2D) atomic-layer systems, such as monolayered hexagonal BN, transition metal dichalcogenides (TMDCs)\cite{D.Xiao,A.K.Geim2,Q.H.Wang}, silicene\cite{P.Vogt,D.Q.Fang}, and germanane\cite{D.Q.Fang,E.Bianco}.

\begin{figure}[b]
  \centering
  \includegraphics[width=0.475\textwidth]{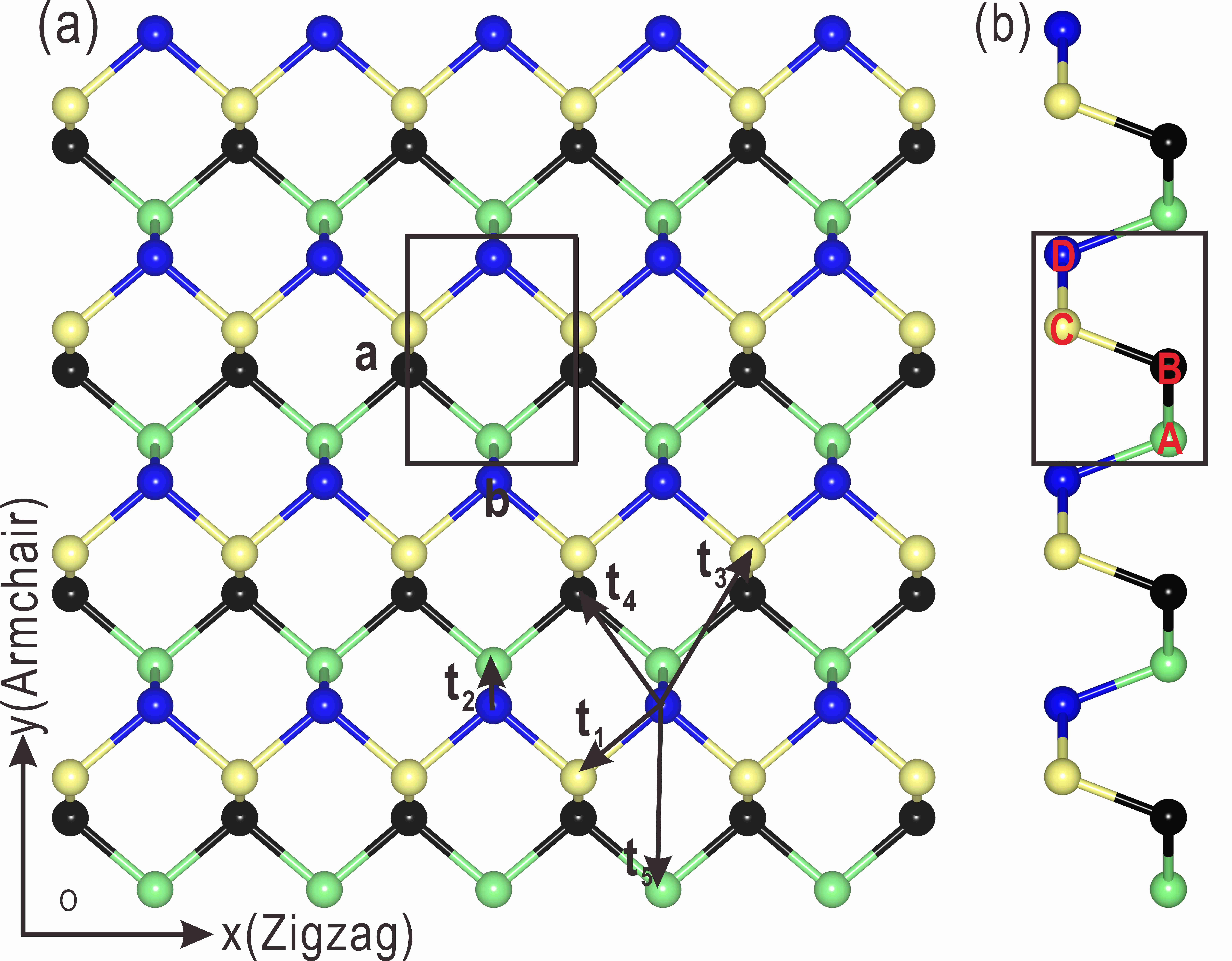}
  \caption{(Color online) (a) Top and (b) side view of phosphorene. The hopping terms in the tight-binding Hamiltonian are shown by edges between sites. There are four phosphorus atoms in a unit cell labeled A, B, C, and D.  }\label{fig:1}
\end{figure}

Recently, a new 2D material, few-layer black phosphorus (termed 'phosphorene'), was successfully fabricated from bulk black phosphorus in experiments\cite{L.Li,H.Liu}. BP consists of phosphorus atom layers, where adjacent layers coupled by weak Van der Waals interactions\cite{A.Morita}. Inside a single layer, each phosphorus atom is covalently bonded with three nearest phosphorus atoms to form a puckered honeycomb structure (Fig.~\ref{fig:1}(a)). BP possesses a direct band gap (0.3 eV) located at Z point. This direct gap increases to 1.5-2 eV when the film thickness decreases from bulk to few layers and eventually monolayer via mechanical exfoliation. This band gap is crucial for practical applications in field-effect transistors\cite{L.Li}. Unlike other widely studied 2D materials such as graphene and TMDCs, the band structure, electrical conductivity, thermal conductivity, and optical responses of few-layer BP are all highly anisotropic\cite{A.S.Rodin,H.Liu,V.Tran,R.Fei,J.Qiao,R.Fei1}.

As a newly emerged member of the 2D crystal family, phosphorenes have recently gained significant attention due to their interesting electrical and optical properties. To date, various properties of phosphorene have been predicted theoretically and verified experimentally, including those related to strain induced gap modification\cite{A.S.Rodin}, field transistor effect\cite{L.Li}, anisotropic Landau levels\cite{X.Y.Zhou}, quantum oscillations\cite{LikaiLi,XiaolongChen,V.Tayari}, and quantum Hall effect\cite{LikaiLi}. An important aspect on the applications of phsophorene is the finite size effect, which has been explored mainly in  phosphorene nanoribbons\cite{M.Ezawa,W.Li,J.Zhang,Taghizadeh}. Phosphorene quantum dots have not yet been researched extensively so far.

Motivated by the recently experimental sysnethis of few-layer phosphorene quantum dots by chemical method\cite{X.Zhang},
we investigate the electronic and magneto-optical properties of monolayer phosphorene quantum dots (MPQDs) here. The electronic structure of monolayer phosphorene (MLP) can be well described by a tight-binding (TB) model\cite{A.N.Rudenko}. Five types of MPQDs with a rectangular, hexangular, and triangular shape are considered. We find edge states detached from the bulk band gap in all kinds of MPQDs studied regardless of their shapes and edge configurations due to the anisotropic hoppings parameters in MLP. The magnetic levels of MPQDs exhibit a Hofstadter-butterfly spectrum and approach the Landau levels of MLP as the magnetic field increases. A "flat band" appears in the magneto-energy spectrum which is totally different from that of MLP. The electronic and optical properties can be tuned by the size and edge type of MPQDs and the external magnetic field.

The paper is organized as follows. In Sec. \Rmnum{2}, we present the TB model. In Sec. \Rmnum{3}, we calculate the electronic properties including the density of state (DOS), probability distribution of edge states, and size dependent energy gap of MPQDs. In  Sec. \Rmnum{4}, we investigate the magneto-energy spectrum and magneto-optical spectrum.
Finally, we summarize our results in Sec. \Rmnum{5}.

\section{TIGHT-BINDING MODEL}
As shown in Fig.~\ref{fig:1}(a) , MLP has an irregular honeycomb structure with lattice constants  $a$=4.38{\AA} and $b$=3.31{\AA}. There are four types of  phosphorus atoms in a unit cell (Fig.~\ref{fig:1}(b)). The Hamiltonian for MPQDs can be written as\cite{A.N.Rudenko}
\begin{equation}\label{(1)}
H=\sum\limits_{i\neq
j}t_{i,j}c_{i}^{\dagger }c_{j} ,
\end{equation}
where the summation runs over all the lattice sites of MPQDs, $c_i^\dag$ ($c_j$) is the creation (annihilation) operator of the electron at site $i$ ($j$), and ${t_{i,j}}$ are the hopping energies. Five hopping links (see Fig.~\ref{fig:1}(a)) are needed to be taken into consideration\cite{A.N.Rudenko}. The related hopping integrals are $t_{1}$=$-$1.220 eV, $t_{2}$=3.655 eV, $t_{3}$=$-$0.205 eV, $t_{4}$=$-$0.105 eV, and $t_{5}$=$-$0.055 eV. The band gap of MLP given by this TB model is 1.52 eV with the valence band maximum (VBM) and a conduction band minimum (CBM) located at $-$1.18 eV and 0.34 eV
respectively\cite{X.Y.Zhou}. When we consider a magnetic field $B$ applied perpendicularly to the plane of a MPQD, the transfer integral becomes ${\widetilde{t}_{i,j}}$ = ${t_{i,j}}{e^{i2\pi {\phi _{i,j}}}}$, where $\phi_{i,j}=\frac{e}{h}\int\nolimits_{r_{i}}^{r_{j}}d\mathbf{l\cdot A}$ is the Peierls phase. In our calculation, we take the Landau gauge ${\mathbf{A}=(0,Bx,0)}$ . The magnetic flux $S=\frac{Bab}{2}$  through a plaquette is in unit of $\Phi_{0}=\frac{e}{h}$. The eigenvalues and eigenstates can be obtained from the secular equation det$|EI-H|$=0, where $H_{i,j}$=$t_{i,j}{e^{i2\pi {\phi _{i,j}}}}$.

The electronic density of states (DOS) of a quantum dot (QD) are the sum of a series of Dirac $\delta$ functions.
Instead of plotting the level structure it is more useful to study the density of levels since it manifests more clearly
the exact and nearly exact degeneracies of levels. Further, a regular variation of the level density will indicate a shell structure. The energy DOS with a Gaussian broadening of a  set of discrete levels $\epsilon_{i}$ is given by
\begin{equation}\label{2}
D(\epsilon)=\frac{1}{\Gamma\sqrt{\pi}}\sum\limits_{i}e^{-(\epsilon-\epsilon_{i})^2/\Gamma^2}
\end{equation}
where $ \Gamma$ is the broadening factor, which is taken as 0.1 eV in our calculations without specification.

\begin{figure}[b]
  \centering
  \includegraphics[width=0.475\textwidth]{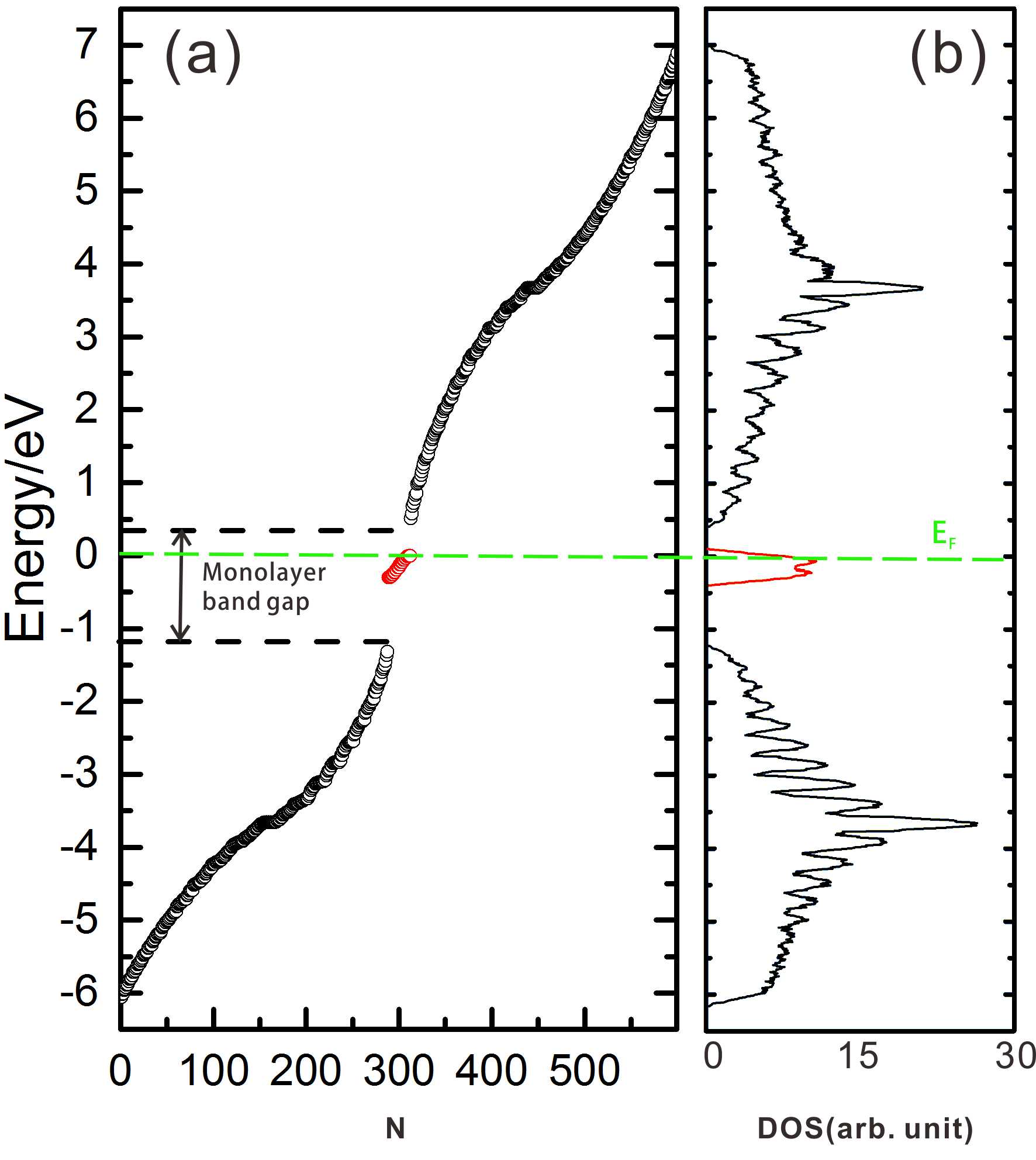}
  \caption{(Color online) (a) Eigenenergies of the RMPQD with $N_{a}=N_{z}=12$.
  The $x$ axis numbers the states with increasing energy. (b) DOS of the RMPQD. We use a Gaussian function with a broadening factor $ \Gamma$=0.1eV
  to smooth the discontinuous energy spectra. The edge states are marked by red color. The horizontal green line is the Fermi level of this dot.}\label{fig:2}
\end{figure}
\section{Electronic properties of MPQDs}
In this paper, five types of MPQDs will be explored including rectangular MPQDs (RMPQDs), hexangular armchair MPQDs (HAMPQDs), hexangular zigzag MPQDs (HAMPQDs), triangular armchair MPQDs (TAMPQDs) and triangular zigzag MPQDs (TZMPQDs). The basic parameters of MPQDs and the information of five typical MPQDs are shown in TABLE.~\ref{tabular:1}. The total number of phosphorous atoms in a MPQD, $N_{total}$, is determined by the number of hexagonal units along the armchair and/or zigzag direction ($N_{a}$ and/or $N_{z}$). The sizes adopted in numerical calculations of five typical MPQDs are comparable to that in experiments\cite{X.Zhang}.
\begin{table}[h]%
\begin{tabular*}
{0.5\textwidth}[c]{@{\extracolsep{\fill} }rclc}\hline\hline
& $Index(N)$ & $N_{total}$& $Calculated[N;Size]$ \\\hline
$\mathrm{RMPQD}$ & $N_{a}$,$N_{z}$ & $(4N_{a}+2)N_{z}$ & $N_{a}$=12,$N_{z}$=12; 40${\AA}$ \\
$\mathrm{HAMPQD}$ & $N_{a}$ & $6(3N_{a}^2-3N_{a}+1)$ &$N_{a}$=7; 56${\AA}$\\
$\mathrm{HZMPQD}$ & $N_{z}$ & $6N_{z}^2$ & $N_{z}$=12; 52${\AA}$ \\
$\mathrm{TAMPQD}$ & $N_{a}$ & $3(N_{a}^2+N_{a})$  & $N_{a}$=16; 69${\AA}$ \\
$\mathrm{TZMPQD}$ & $N_{z}$ & $(N_{z}+1)^2-3$ & $N_{z}$=20; 65${\AA}$ \\\hline\hline
\end{tabular*}
\caption{Basic information of MPQDs including the index, number of phosphorus atoms $N_{total}$ and the size of MPQDs studied. $N_{a}$ and $N_{z}$ are the number of hexagonal units along armchair and zigzag edge. RMPQDs are characterized by two indexs $N_{a}$ and $N_{z}$, while other MPQDs are characterized by $N_{a}$ or $N_{z}$. }%
\label{tabular:1}%
\end{table}
\subsection{Rectangular MPQDs}
We begin with the electron structures in RMPQDs under zero magnetic field. The discrete levels and smoothing energy DOS
for a RMPQD with $N_{a}=N_{z}=12$ are presented in Fig.~\ref{fig:2}. Edge states appear in the band gap of MLP (mid-in-gap states) below Fermi level ($E_{F}$) (red point in Fig.~\ref{fig:2}(a)). The corresponding DOS posses a peak within the MLP band gap (see Fig.~\ref{fig:2}(b)), which is below Fermi level ($E_{F}$) and marked by red color. The edge states play important roles in the electronic and magneto properties of RMPQDs, which makes the RMPQDs different to MLP.
\begin{figure}[h]
  \centering
  \includegraphics[width=0.475\textwidth]{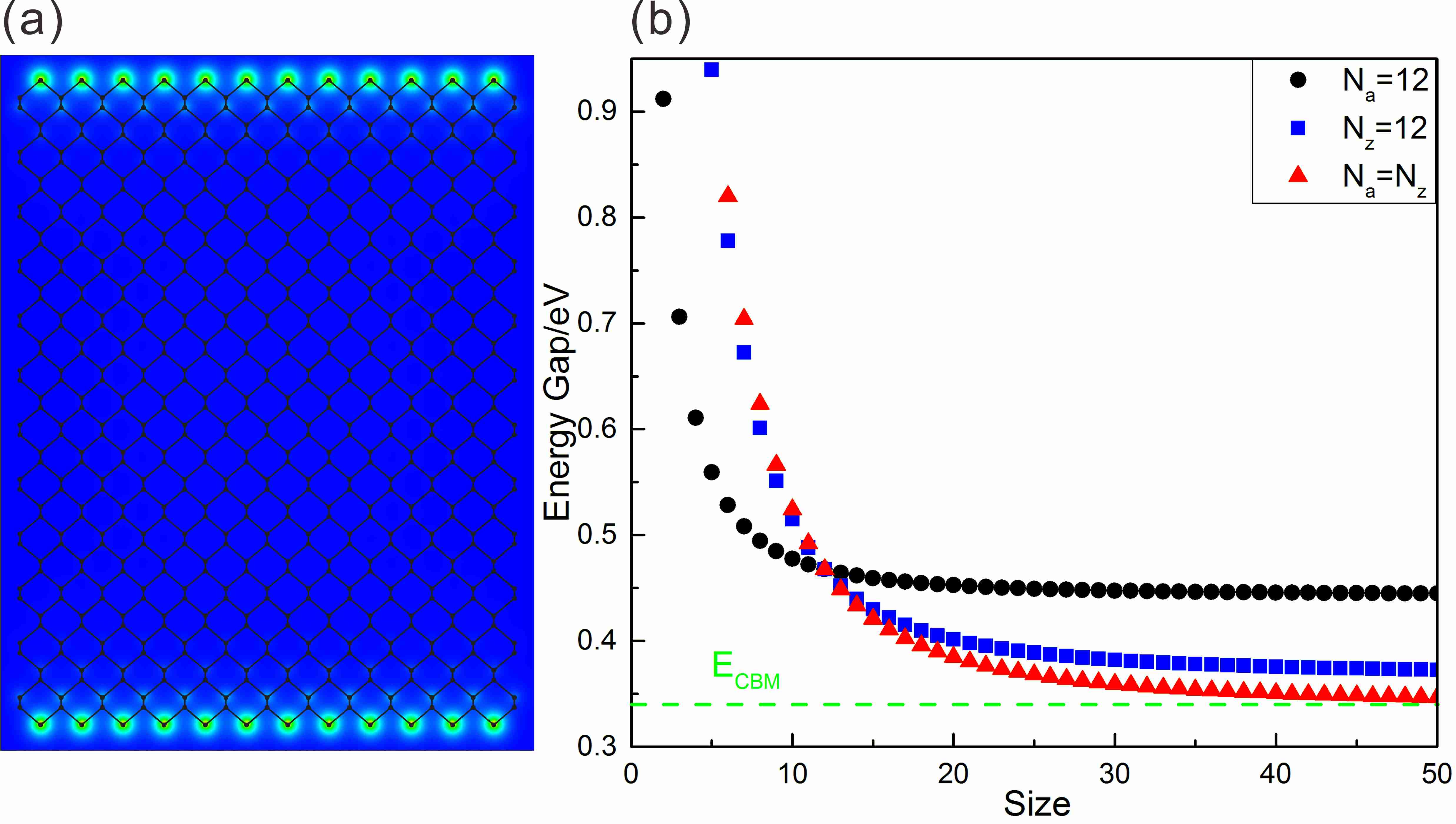}
  \caption{(Color online) (a) The electronic density distributions of mid-in-gap states of RMPQD with $N_{a}=N_{z}=12$.
  (b) Energy gap as a function of the size of RMPQDs. We change the dot size in three ways: (1) increase $N_{z}$ but fix $N_{a}$=12; (2) increase $N_{a}$ but fix $N_{z}$=12; (3) keep $N_{a}$=$N_{z}$ and increase them. $E_{CBM}$=0.34 eV
  is the conduction band minimum of monolayer phosphorene. }\label{fig:3}
\end{figure}
The probability distributions contributed by all edge states in the RMPQD with $N_{a}=N_{z}=12$ is shown in Fig.~\ref{fig:3}(a). We find that electrons in all edge states localize at the phosphorus atoms in the zigzag edge. Further , the number of mid-in-gap states is exactly equal to the number of phosphorus atoms in the zigzag edges in RMPQDs. This observation is important in tuning the electronic and magneto-optical properties of RMPQDs by changing the dot size.

In Fig.~\ref{fig:3}(b), we examine the effect of the dot size (the parameters $N_{a}$ and $N_{z}$) on the energy gap for RMPQDs. We change the dot size in three ways: (1) increase $N_{z}$ but fix $N_{a}$=12; (2) increase $N_{a}$ but fix $N_{z}$=12; (3) keep $N_{a}=N_{z}$ and increase them. In all three cases, the energy of the lowest unoccupied molecular orbital (LUMO)  decreases with the dot size increasing, the energy of the highest occupied molecular orbital (HOMO) keeps constant (approximated by $E_{F}$).  Therefore the gap between HOMO and LUMO does not approach the energy gap of MLP (1.52 eV). When $N_{a}=N_{z}$, the energy of LUMO will decrease to the CBM of 2D phosphorene. In the other two cases where either $N_{a}$  or $N_{z}$ is fixed, the energy of LUMO  decreases to a limit value higher than $E_{CBM}$ of MLP, but approaches the CBM of corresponding ribbon with a fixed width. When we increase the size by method (1), the LUMO and HOMO of the RMPQDs will approach the CBM and VBM of zigzag-edged monolayer phosphorene ribbons (ZMPR) with a width $N=12$ respectively. The reason is that ZMPR poss topological protected edge states\cite{M.Ezawa} . When we increase the size by method (2), only the LUMO of RMPQDs approaches the CBM of armchair-edged monolayer phosphorene ribbons (AMPR) with a width $N=12$, while the HOMO of RMPQDs is higher than the VBM of AMPR with a width $N=12$, because there are no edge states in the AMPR.

\subsection{Hexagonal and triangular MPQDs}

\begin{figure}[b]
  \centering
  \includegraphics[width=0.475\textwidth]{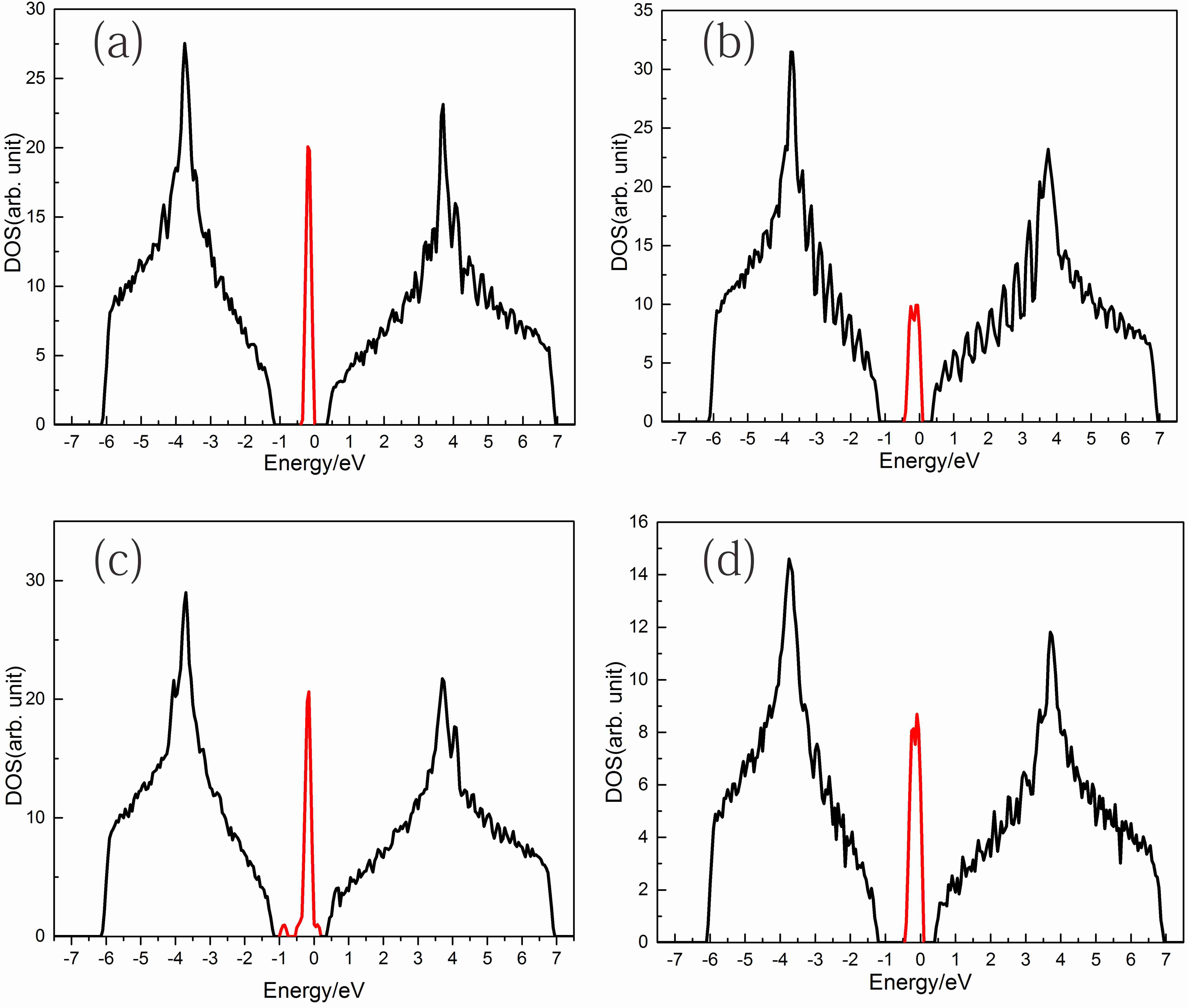}
  \caption{(Color online) Density of state of  MPQDs. (a) HAMPQD, $N_{a}=7$.  (b) HZMPQD, $N_{z}=12$. (c) TAMPQD, $N_{a}=16$. (d) TZMPQD, $N_{z}=20$.
  The region where mid-gap states exist is marked by red color.}\label{fig:4}
\end{figure}
\begin{figure}[t]
  \centering
  \includegraphics[width=0.475\textwidth]{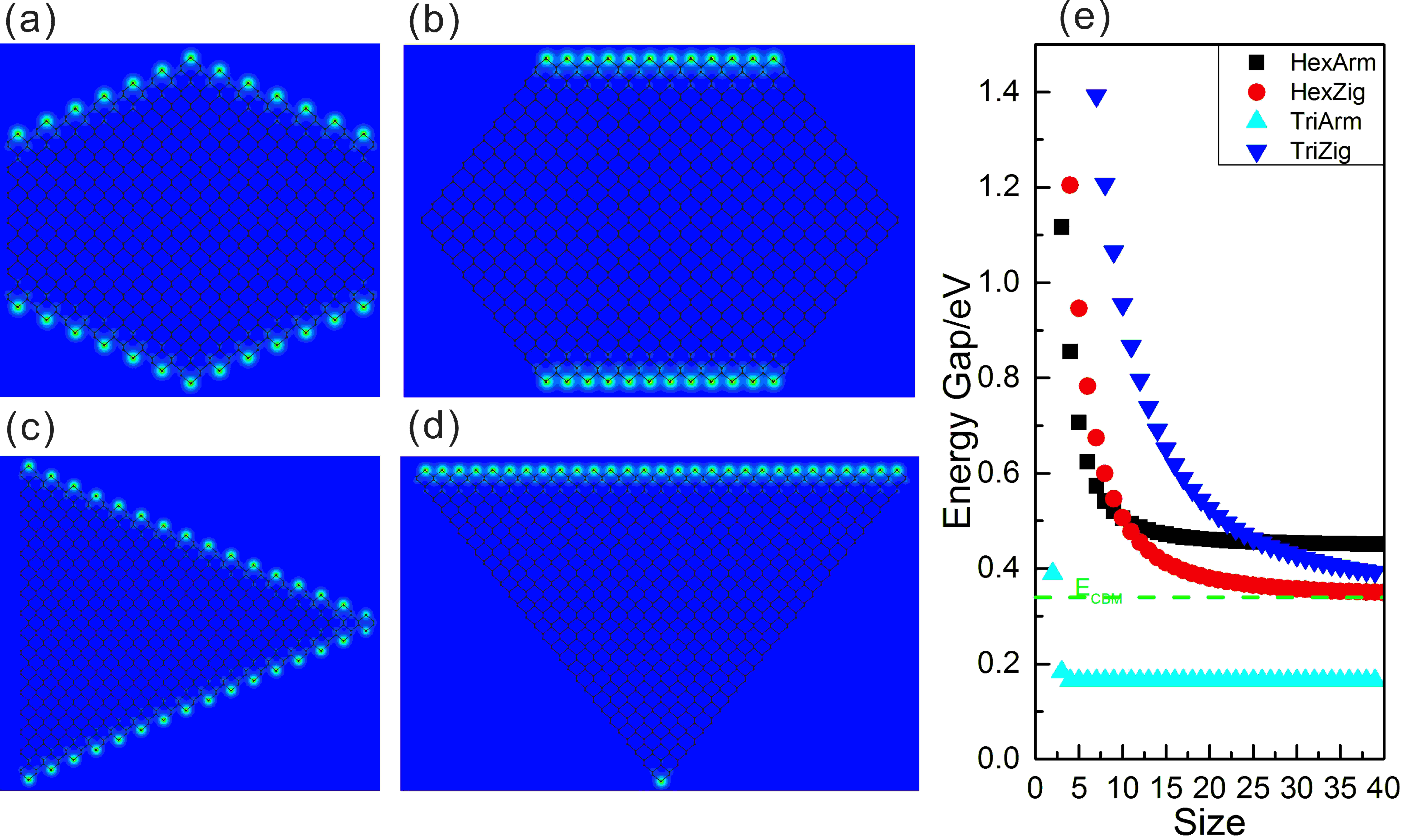}
  \caption{(Color online) Electronic density distributions of edge states. (a)HAMPQD with $N_{a}=7$.
  (b) HZMPQD with $N_{z}=12$. (c) TAMPQD with $N_{a}=16$. (d) TZMPQD with $N_{z}=20$. (e) Energy gap as a function of size of MPQDs.}\label{fig:5}
\end{figure}
Now we explore the electronic properties of HAMPQDs, HZMPQDs, TAMPQDs, and TZMPQDs. For the four kinds of MPQDs,
the numbers of hexagonal units ($N_{a}$ or $N_{z}$) along  all edges of their boundary are identical.

In FIG.~\ref{fig:4}, we plot the DOS spectrum of electrons in hexagonal and triangular MPQDs with different edge shapes.
It is evident that mid-gap states appear near the Fermi energy $E_{F}$ (see the red curves), regardless of the boundary shape or the edge types. The general existence of edge states in MPQDs is remarkable, which is different to that of graphene quantum dots (GQDs). In GQDs, edge states only appear in hexagonal zigzag edged GQDs \cite{Z.Z.Zhang,AlevDevrim}.

The probability distributions of edge states in these MPQDs are shown in Fig.~\ref{fig:5}. For all edge states considered here electrons appear  only at $\textbf{A-type}$ and $\textbf{D-type}$ (see Fig.~\ref{fig:1}(b)) terminated phosphorus atoms in the armchair direction, which is distinct to that in GQDs. In GQDs, the edge states are type-selected, i.e., the edges of a GQD are identical and the edge states appear or disappear in every edge of a GQD. However, edge states in hexagonal and triangular MPQDs are not type-selected but direction-selected, that's to say the electrons in edge states will appear at $\textbf{A-type}$ and $\textbf{D-type}$ terminated phosphorus atoms in the armchair direction , but disappear at $\textbf{B-type}$ and $\textbf{C-type}$ terminated phosphorus atoms. This conclusion is also suitable to RMPQDs. To illustrate this difference, the easiest way is to find out the difference in the TB hamiltonian. In the GQDs, the hopping to three nearest neighbor carbon atoms are identical, which lead to the isotropic properties of GQDs. In  MPQDs the hopping integral to the nearest three phosphorus are $t_{1}$ (hopping along zigzag direction) and $t_{2}$ (hopping along armchair direction). The large difference between $t_{1}$ and $t_{2}$ makes MPQDs highly anisotropic. This is also the main reason why MPQDs have direction-selected edge state.

The variation of the energy gap with the size of MPQDs is plotted in  Fig.~\ref{fig:5}(e). The energy gap decreases as the size of the MPQD increases except for TAMPQDs. The energy of HOMO and LUMO of large size MPQDs are shown in TABLE.~\ref{tabular:2}. From TABLE.~\ref{tabular:2}, we can see that the HOMO of this four types of MPQD are from the edge states, while the LUMO are from the bulk state except for TAMPQDs.  In TAMPQDs, both HOMO and LUMO are  from the edge states. Therefore, for TAMPQDs the energy gap keeps constant as the dot size increases.
\begin{table}[h]%
\begin{tabular*}
{0.48\textwidth}[c]{@{\extracolsep{\fill} }rccc}\hline\hline
& HOMO(eV) & LUMO(eV)& Gap(eV) \\\hline
$\mathrm{HAMPQD}$ & -0.11 & 0.34 &0.45\\
$\mathrm{HZMPQD}$ & 0 & 0.34 & 0.34 \\
$\mathrm{TAMPQD}$ & -0.07 &  0.10  & 0.17 \\
$\mathrm{TZMPQD}$ & 0 & 0.35 & 0.35 \\\hline\hline
\end{tabular*}
\caption{The energy of HOMO, LUMO and the gap between HOMO and LUMO when we increase $N_{a}$ (or $N_{z}$) to 50. }%
\label{tabular:2}%
\end{table}
\section{MPQDS UNDER MAGNETIC FIELDS}
To reveal potential applications of MPQDs, it is necessary to study the magneto-energy spectrum and optical absorption.
In this section we take the RMPQD with $N_{a}$=$N_{a}$=12 as an example to examine the effect of magnetic
fields on both the energy spectrum and optical absorption.

\begin{figure}[t]
  \centering
  \includegraphics[width=0.475\textwidth]{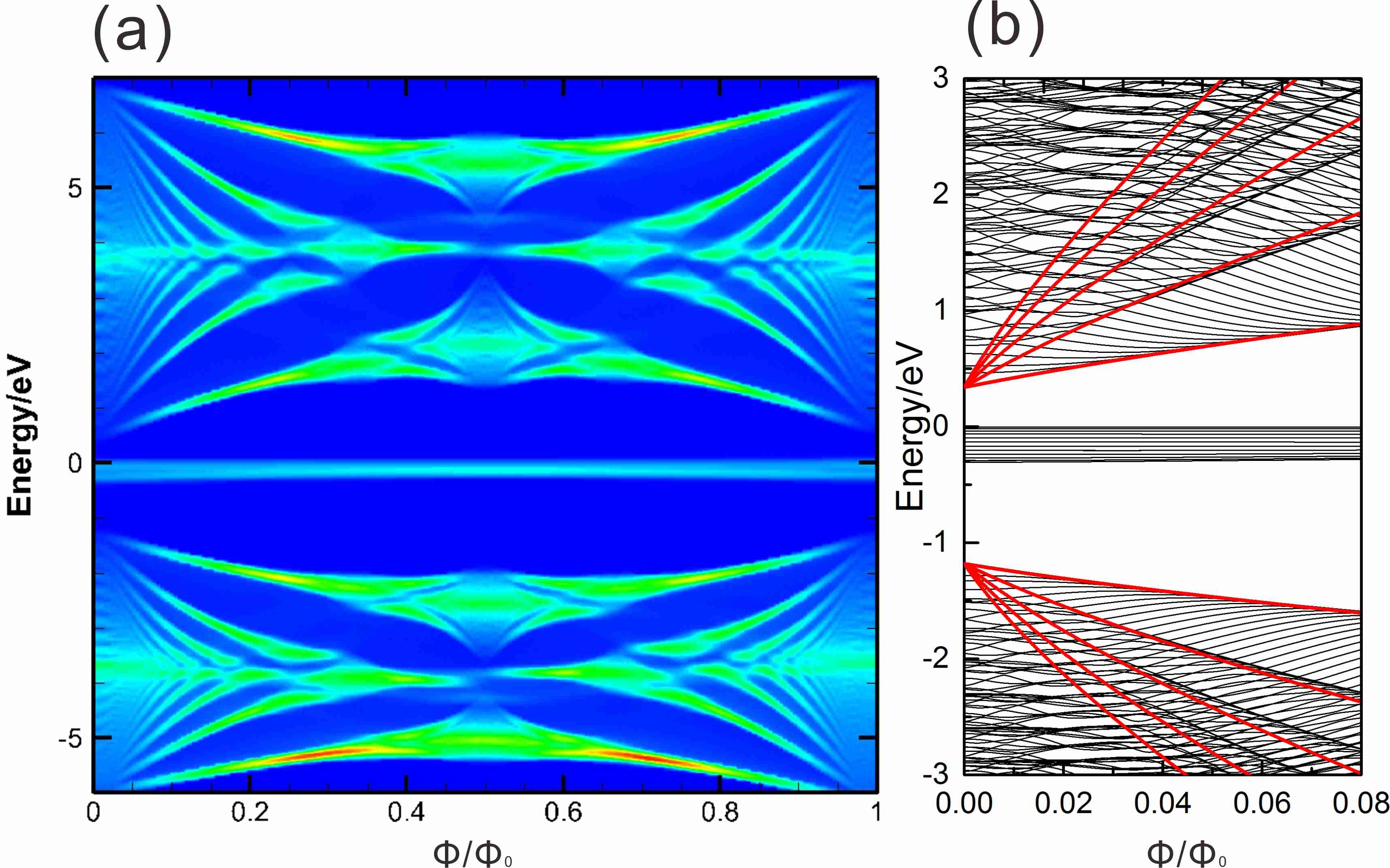}
  \caption{(Color online) (a) The DOS and energy spectrum of the RMPQD with $N_{a}=N_{z}$=12 in the presence of magnetic field.
  We use a Gaussian function with broadening factor 0.01 eV here. (b) The magnetic energy spectra in relatively
  low magnetic field. The red lines correspond to the Landau levels of MLP.  }\label{fig:6}
\end{figure}

The calculated magneto-energy spectrum is shown in Fig.~\ref{fig:6}(a), which exhibits a clear Hofstardter-butterfly feature. As the magnetic flux increases, the magnetic levels in the RMPQD approach the Landau levels of MLP\cite{X.Y.Zhou}, as shown in Fig.~\ref{fig:6}(b). A "flat band" appears near the Fermi energy $E_{F}$, which is derived from the edge states at zero magnetic field. This is because the edge states have a negligible probability distribution with a distance smaller than or comparable to the magneto length ($l_{B}=\sqrt{\hbar/eB}$).

\begin{figure}[h]
  \centering
  \includegraphics[width=0.475\textwidth]{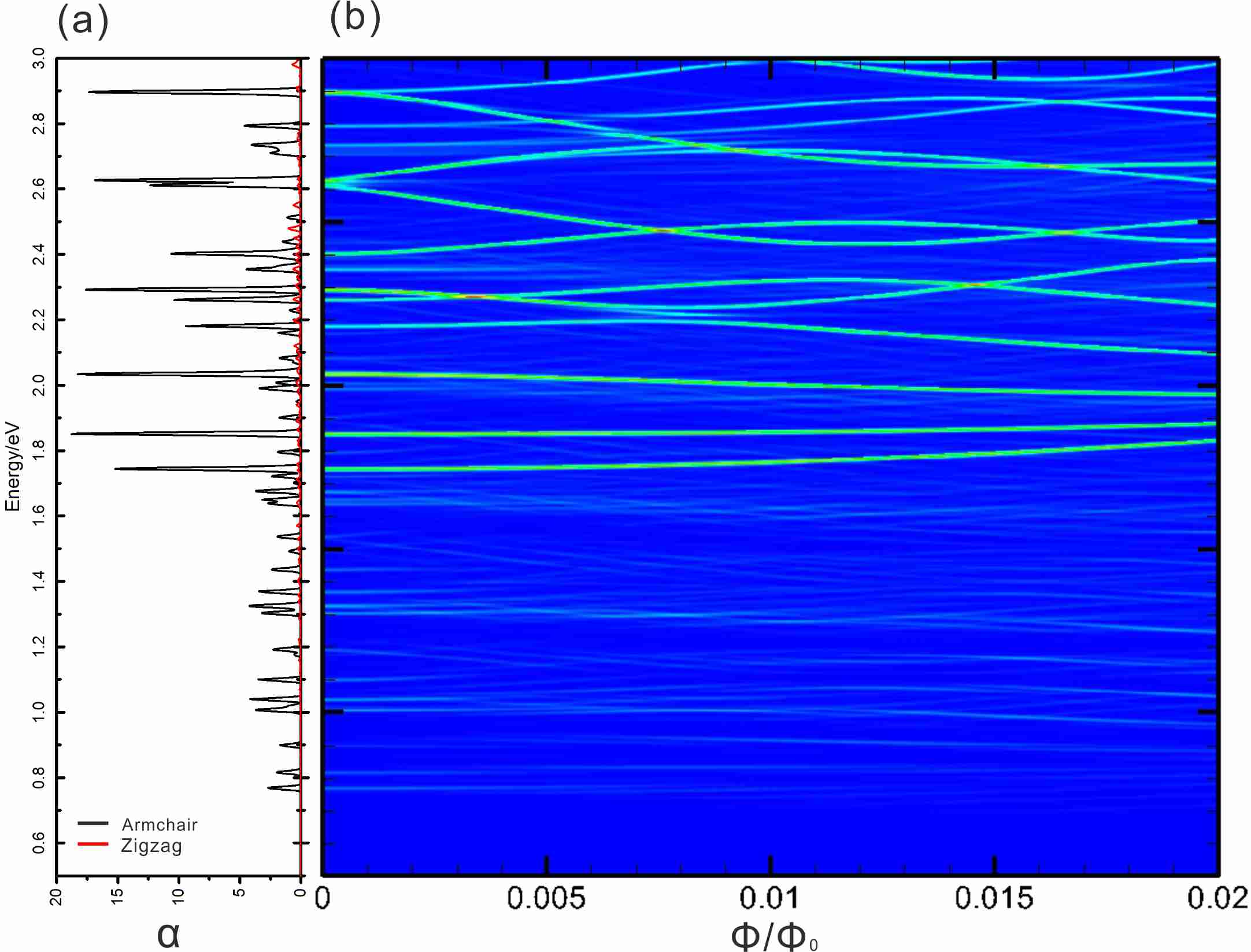}
  \caption{(Color online)  (a) The optical-absorption spectrum $\alpha$ for RMPQD ($N_{a}=N_{z}=12$) in the absence of magnetic field.
  The black and red lines represent the absorption of light with polarization along the armchair and zigzag direction.
  We use a Gaussian function with broadening factor 0.005 eV to smoothen the discontinuous absorption spectra.
  (b) The contour plot of the magneto-optical spectra of RMPQD ($N_{a}=N_{z}=12$). }\label{fig:7}
\end{figure}
The optical properties of MPQDs are promising for potential applications in optic-electronic devices based on
phosphorene. Therefore, we calculate the absorption spectra of RMPQD with $N_{a}=N_{z}$=12,
$$\alpha (\hbar \omega )=%
\frac{\pi e^{2}}{m_{0}^{2}\varepsilon _{0}cn\omega V}\sum\limits_{c,v}\left%
\vert \overrightarrow{\varepsilon }\cdot P_{cv}\right\vert ^{2}\times \delta
(E_{c}-E_{v}-\hbar \omega )$$
where $n$ is the refractive index, $c$ is the speed of light in vacuum, ${\varepsilon _{0}}$
the permittivity of vacuum, $m_{0}$ is the free-electron mass, and $\overrightarrow{\varepsilon }$ is
the polarization vector of the incident light. The moment matrix is
$$\left\langle n\right\vert p\left\vert m\right\rangle =im_{0}/\hbar\sum\limits_{r}\sum\limits_{r^{\prime }}c_{n,r}^{\ast }
c_{m,r^{\prime}}(r^{\prime }-r)\left\langle r^{\prime }\right\vert H\left\vert
r\right\rangle $$
here $\left\langle r^{\prime }\right\vert H\left\vert r\right\rangle $ is the hopping integral between $r$ and $r^{\prime}$.

The optical-absorption spectrum for RMPQD with $N_{a}=N_{z}=12$ in the absence of magnetic field is shown in Fig.~\ref{fig:7}(a). The absorption of incident light with polarization along the armchair direction is about 20 times stronger than the absorption  of incident light with polarization along the zigzag direction. The result is qualitatively the same to that of MLP\cite{S.Yuan}. For incident light with polarization along the armchair direction, the absorption peak below 1.52 eV results from the transition from the edge states to the bulk states, which would be important in future applications of MPQDs. Note that the intensity of these absorption peaks are relatively weaker than those absorptions due to transtion  between bulk states. The main reason is that the phosphorus atoms on the MPQD boundary are much less than bulk phosphorus atoms. The effect of a magnetic field on the optical spectrum of the RMPQD is shown in Fig.~\ref{fig:7}(b). In the contour magneto-optical graph, the strong absorption lines exhibit asymptotic behavior corresponding to the transitions between the conduction and valence band Landau levels at a high magnetic field.

\section{SUMMARY}
We have theoretically investigated the electronic and magneto-optical properties of rectangular,
hexangular and triangular MPQDs by means of the TB model. We find edge states detached from the bulk band gap in all kinds of MPQDs regardless of their shapes and edge configurations due to the anisotropic hoppings parameters in MLP. For all edge states, the electron density is concentrated only in the armchair direction of the dot boundary, which is distinct to that in graphene quantum dots. The magnetic levels of MPQDs exhibit a Hofstadter-butterfly spectrum and approach the Landau levels of MLP as the magnetic field increases. A "flat band" appears in the magneto-energy spectrum which is totally different from that of MLP. The electronic properties and optical absorption of MPQDs can be tuned by changing the dot size, the types of boundary edges, and the external magnetic field.

\section{ACKNOWLEDGMENTS}

This work was supported by the NSFC Grants No.11434010, No.2011CB922204-2, NO.11174252, 2015CB921503, and 2012CB934304 from the MOST of China.

\bibliographystyle{apsrev4-1}

\end{document}